\documentclass[sigconf]{acmart}
\AtBeginDocument{%
  \providecommand\BibTeX{{%
    \normalfont B\kern-0.5em{\scshape i\kern-0.25em b}\kern-0.8em\TeX}}}

\usepackage{hyperref}
\usepackage{listings}
\usepackage{adjustbox}
\usepackage{geometry}
\usepackage{makecell}
\usepackage{multirow}
\usepackage{enumitem}
\usepackage{tabularx}
\usepackage{array}
\usepackage{graphicx}
\usepackage{longtable}

\title{TAMIGO: Empowering Teaching Assistants using LLM-assisted viva and code assessment in an Advanced Computing Class}

\author{Anishka}
\email{anishka20282@iiitd.ac.in}
\affiliation{%
  \institution{IIIT Delhi}
  \state{Delhi}
  \country{India}
}

\author{Diksha Sethi}
\email{diksha20056@iiitd.ac.in}
\affiliation{%
  \institution{IIIT Delhi}
  \state{Delhi}
  \country{India}
}

\author{Nipun Gupta}
\email{nipun20089@iiitd.ac.in}
\affiliation{%
  \institution{IIIT Delhi}
  \state{Delhi}
  \country{India}
}

\author{Shikhar Sharma}
\email{shikhar20121@iiitd.ac.in}
\affiliation{%
  \institution{IIIT Delhi}
  \state{Delhi}
  \country{India}
}

\author{Srishti	Jain}
\email{srishti20543@iiitd.ac.in}
\affiliation{%
  \institution{IIIT Delhi}
  \state{Delhi}
  \country{India}
}

\author{Ujjwal Singhal}
\email{ujjwal21434@iiitd.ac.in}
\affiliation{%
  \institution{IIIT Delhi}
  \state{Delhi}
  \country{India}
}

\author{Dhruv Kumar}
\email{dhruv.kumar@iiitd.ac.in}
\affiliation{%
  \institution{IIIT Delhi and BITS Pilani}
  \state{Delhi}
  \country{India}
}

\begin{document}



\renewcommand{\shortauthors}{Anony Mous et al}

\begin{abstract}

Large Language Models (LLMs) have significantly transformed the educational landscape, offering new tools for students, instructors, and teaching assistants. This paper investigates the application of LLMs in assisting teaching assistants (TAs) with viva and code assessments in an advanced computing class on distributed systems in an Indian University. We develop TAMIGO, an LLM-based system for TAs to evaluate programming assignments.

For viva assessment, the TAs generated questions using TAMIGO and circulated these questions to the students for answering. The TAs then used TAMIGO to generate feedback on student answers. For code assessment, the TAs selected specific code blocks from student code submissions and fed it to TAMIGO to generate feedback for these code blocks. The TAMIGO-generated feedback for student answers and code blocks was used by the TAs for further evaluation.

We evaluate the quality of LLM-generated viva questions, model answers, feedback on viva answers, and feedback on student code submissions. Our results indicate that LLMs are highly effective at generating viva questions when provided with sufficient context and background information. However, the results for LLM-generated feedback on viva answers were mixed; instances of hallucination occasionally reduced the accuracy of feedback. Despite this, the feedback was consistent, constructive, comprehensive, balanced, and did not overwhelm the TAs. Similarly, for code submissions, the LLM-generated feedback was constructive, comprehensive and balanced, though there was room for improvement in aligning the feedback with the instructor-provided rubric for code evaluation. Our findings contribute to understanding the benefits and limitations of integrating LLMs into educational settings.

\end{abstract}




\received{20 February 2007}
\received[revised]{12 March 2009}
\received[accepted]{5 June 2009}

\maketitle

\section{Introduction}
In recent years, the advent of Large Language Models (LLMs) has significantly influenced various domains, including education. Models, such as OpenAI's GPT-3.5 and GPT-4 \cite{openai_gpt-4_2023}, Google's Gemini \cite{geminiteam2024geminifamilyhighlycapable}, Meta's Llama-3 \cite{Meta}, possess the capability to generate human-like text and engage in critical and logical thinking, making them valuable tools for a variety of educational tasks. LLMs can aid in various applications ranging from aiding students in their learning processes to supporting instructors and teaching assistants (TAs) in the evaluation and assessment of student work. Since the LLM-based technology is rapidly evolving, it becomes imperative to explore and understand the potential and limitations of integrating such advanced technology into academic settings.

Existing research on the integration of LLMs in computing education has explored the challenges and opportunities that educators and students encounter when adapting to LLMs' capabilities, particularly in generating code from natural language descriptions \cite{becker2023ProsAndCons, Malinka2023Security, Daun2023Software, Denny2023CopilotCS1, Finnie-Ansley2022CS1, wermelinger2023Copilot, Savelka2023MCQAndCode, Reeves2023Parsons, finnie-ansley2023CodexCS2, Ouh2023Java, Cipriano2023GPT-3OOP, sarsa2022AutoGenerate, Leinonen2023CodeExplanation, Leinonen2023ExplainError, MacNeil2023CodeExplain, Balse2023Feedback, joshi2023chatgpt, joshi2023interviews}. The primary focus has been on introductory programming courses, with applications including code generation \cite{Denny2023CopilotCS1, Finnie-Ansley2022CS1, wermelinger2023Copilot, Savelka2023MCQAndCode, Reeves2023Parsons, finnie-ansley2023CodexCS2}, code explanation \cite{MacNeil2023CodeExplain, Leinonen2023CodeExplanation}, debugging \cite{Leinonen2023ExplainError}, and the development of supportive tools for students \cite{kazemitabaar2024codeaid, codehelpusinglarge}. These studies have demonstrated the potential of LLMs to enhance the educational experience by providing tailored assistance and feedback, thereby aiding students' in thier learning processes instructors in their teaching efforts.


This paper presents an experience report into the application of LLMs in assisting the teaching assistants (TAs) with viva and code assessments for take-home programming assignments in an advanced computing class on distributed systems at an Indian University. Similar to global scenarios, the TAs in Indian universities are responsible for a multitude of tasks, including conducting tutorials and labs, resolving queries, conducting viva assessments, grading programming assignments, and providing constructive feedback to students. However, these tasks are often time-consuming and require a high level of expertise. This is where LLMs can play a crucial role by augmenting the capabilities of TAs, thereby enhancing the efficiency and effectiveness of the evaluation process. 

Exploring the feasibility and effectiveness of an LLM-based system for viva and code evaluation was driven by several factors. Firstly, the preparation of questions for student evaluations, particularly for vivas in take-home programming assignments, is time-consuming and labor-intensive. Since vivas are conducted by TAs at different times, using a shared list of questions risks leaking these questions to students who have not yet been assessed. LLMs can generate diverse and contextually relevant questions, reducing this risk. Secondly, TAs often have varying levels of subject matter expertise and grading strictness, leading to inconsistencies in evaluation. LLM-based tools can contribute to more consistent and fair assessments by providing standardized feedback and evaluation criteria. Lastly, the evaluation of programming assignments in advanced computing courses involves extensive review of student-written code, which can be a daunting and time-consuming task for TAs. LLMs can generate concise code summaries and constructive feedback, significantly saving time and effort for TAs and ensuring a thorough understanding of student submissions.

 We developed TAMIGO, an LLM-based system leveraging OpenAI's API (GPT-3.5-Turbo model), to support TAs in generating viva questions, providing feedback on student answers, and assessing student code submissions. We deployed TAMIGO for the evaluation of two take-home programming assignments in the advanced computing class. For the viva assessment, TAs utilized TAMIGO to generate multiple questions and subsequently used it to generate feedback on student answers. Similarly, for code assessments, TAs selected specific code blocks from student submissions and used TAMIGO to generate comprehensive feedback. The TAs were asked to share all the LLM-generated questions, answer feedback and code assessments with the research team. The research team then evaluated the data collected to determine the quality of LLM-generated viva questions, feedback on viva answers, and feedback on code submissions, providing insights into the advantages and shortcomings of using LLMs for evaluation purposes.

Our findings suggest - LLMs are highly effective at generating questions with sufficient context, the feedback on viva answers exhibited occasional hallucinations, impacting accuracy. Despite this, the feedback was generally constructive and balanced, aiding TAs without overwhelming them. Similarly, the feedback on code submissions was comprehensive, though improvements are needed to better match the course rubric. This study contributes to the broader understanding of the integration of LLMs in education, highlighting both the potential benefits and the challenges.

In the context of advanced computing classes, the main objective of our study is to evaluate the \textbf{quality and effectiveness of: (1) LLM-generated question and model answers, (2)  LLM-generated feedback on student answers, and (3) LLM-generated feedback and summary for student-written code.}
\vspace{-1em}

\section{Related Work}
\noindent\textbf{Educator and Student Perspectives on LLMs.}
Within the computing education research community, several studies and reports have focused on LLMs \cite{becker2023ProsAndCons, Malinka2023Security, Daun2023Software, Denny2023CopilotCS1, Finnie-Ansley2022CS1, wermelinger2023Copilot, Savelka2023MCQAndCode, Reeves2023Parsons, finnie-ansley2023CodexCS2, Ouh2023Java, Cipriano2023GPT-3OOP, sarsa2022AutoGenerate, Leinonen2023CodeExplanation, Leinonen2023ExplainError, MacNeil2023CodeExplain, Balse2023Feedback, joshi2023chatgpt, joshi2023interviews, budhiraja2024Jarvis, Lau2023Instructor}. These studies have shed light on the perspectives of both educators and students regarding the use of LLMs in various contexts. 
While instructors can leverage these AI tools to generate new content and exercises for students, as well as create simplified explanations of computer science concepts, students can use these tools to resolve their doubts, learn new topics, practice and prepare for exams. There is also a growing concern that students may also misuse these tools to obtain complete answers to graded assignments. 

\noindent\textbf{Code Generation Using LLMs.}
A significant number of research studies have specifically analyzed the accuracy of LLMs, such as OpenAI Codex, GPT-3, and ChatGPT (GPT-3.5 and GPT-4), in generating solutions for programming assignments across various computer science courses \cite{Denny2023CopilotCS1, Finnie-Ansley2022CS1, wermelinger2023Copilot, Savelka2023MCQAndCode, Reeves2023Parsons, finnie-ansley2023CodexCS2, Savelka2023MCQAndCode, Ouh2023Java, Cipriano2023GPT-3OOP, Daun2023Software, Malinka2023Security}. 
These studies suggest that LLMs are good at generating code, debugging errors and have the potential to assist students in their learning journey and provide valuable support in various programming-related tasks.


\noindent\textbf{LLMs for Supporting TAs and Students.}
Several studies have also analyzed the ability of LLM models to perform functions traditionally performed by teaching assistants (TAs) \cite{FinnieRobots, Hellas_2023, Nilsson1779778, anishka2024chatgptplayroleteaching, sarsa2022AutoGenerate, Balse2023Feedback, tungphung2023international, hicke2023aita, wermelinger2023Copilot, Leinonen2023CodeExplanation, MacNeil2023CodeExplain, kazemitabaar2024codeaid}. These include answering student questions \cite{hicke2023aita}, generating code explanations \cite{sarsa2022AutoGenerate, wermelinger2023Copilot, Leinonen2023CodeExplanation, MacNeil2023CodeExplain}, assisting in problem-solving by offering hints without revealing the full solution \cite{lee2023learning, kazemitabaar2024codeaid, codehelpusinglarge}, evaluating student code \cite{anishka2024chatgptplayroleteaching}, enhancing programming error messages \cite{Leinonen2023ExplainError}, investigating high-precision feedback for programming syntax errors \cite{tungphung2023international}, generating programming exercises \cite{sarsa2022AutoGenerate} and providing detailed and personalized feedback \cite{Balse2023Feedback, tungphung2023international}.

\noindent\textbf{Automated Grading of Programming Assignments.} Existing approaches to automated grading of programming assignments typically involve evaluating student submissions by comparing them to predefined test cases and identifying coding errors \cite{fordentA-bot}. Additionally, some automated grading systems incorporate instructor-provided reference solutions, assessing student work by comparing it against the logical structure of these model codes using semantic analysis \cite{fanCS1-AutomatedGrading, liu_autograde_assignments}. However, these tools have several limitations. Test case-based grading can overlook logical correctness in student code, and reference code-based auto-graders may not account for the diversity of correct solutions. In contrast, our work explores the application of LLMs in generating qualitative feedback to address these limitations. 

\noindent\textbf{How is our study different from existing research on LLMs with Education?} Our work differs from existing studies in the following ways:(1) Most of the existing studies in computing education research have explored the generation and evaluation capabilities of LLMs in the context of introductory programming classes. Our work is focused on evaluating the aforementioned capabilities of LLMs in advanced computing classes such as Distributed Systems.(2) Not many studies have shared the details and experiences related to real-world deployment and integration of these LLM-based tools into TA workflows. As part of this experience report, we share the complete development and deployment experiences of integrating LLM-based tools into existing TA workflows.

\section{Methodology}
\subsection{Class Demographics and Description}
  \label{fig:Overview of Phases}
We deployed TAMIGO, an LLM-based bot in an advanced computing class on Distributed systems at a Tier 1 Indian University. The main aim of TAMIGO was to assist the TAs in evaluating students for take-home assignments given as part of course evaluation. The class comprised a total of 411 (366 males and 45 females) students out of which there were 161 junior-year undergraduate (UG), 232 senior-year UG, 16 graduate Master’s and 2 PhD students. The course was taught by 1 instructor and supported by 15 Teaching Assistants (TAs). This class provided students with the fundamental knowledge of designing and implementing distributed systems. The class evaluation consisted of four quizzes (10\%, 2.5\% each), a mid-semester examination (20\%), an end-semester examination (30\%), and three take-home programming assignments (total 40\% split into 15\% + 15\% + 10\%).

The assignments required students to implement concepts taught in the class, in groups of 1-3 students, resulting in 151 groups. Each assignment had around 15-days completion window, with submissions including code files and documentation in a zip file. The TAs evaluated the assignments through demos, where each group demonstrated their system's correctness and functionality and also answered viva questions posed by the TAs.

Assignment 1 involved developing distributed systems using communication libraries such as gRPC, RabbitMQ, and ZeroMQ. Assignment 2 focused on implementing a modified version of a popular consensus algorithm, Raft. Assignment 3 required developing a MapReduce-like framework for running K-means clustering in a distributed manner. Complete details about the assignment are available on \href{https://github.com/tamigo-research/tamigo}{GitHub} \cite{tamigoRepo}.

\subsection{Development and Deployment of TAMIGO}
Figure \ref{fig:Overview of Phases} provides an overview of all the phases involved in the development and deployment of TAMIGO. The details of each phase are as follows:
\noindent \textbf{Phase 1 - Project Initialization}.
The course instructor discussed the system's scope and objectives with the developers, aiming to develop an LLM-based tool, TAMIGO to assist with the evaluation of take-home programming assignments in the Distributed Systems class. The key objectives established were: generating contextually relevant questions, ensuring evaluation consistency, and streamlining the viva process for TAs.

\noindent \textbf{Phase 2 - Development of TAMIGO(v1)}. We developed two key modules as part of the first version of TAMIGO(v1): (1) Question Generation module and (2) Answer Evaluation module. The Question Generation module generated relevant viva questions and generated model answers using course materials and assignment descriptions, with outputs serialized into JSON format. The Answer Evaluation module provided detailed feedback on student responses by comparing them against model answers, helping TAs to grade students by highlighting areas of proficiency and deficiency. Initially, we used the chat completions API, but due to irrelevant and varied responses, we switched to Retrieval-Augmented Generation (RAG) for more accurate, context-specific questions and evaluations. We utilized the GPT-3.5-Turbo Model.
\noindent\textbf{Phase 3 - Deployment of TAMIGO(v1)}.
By the time, developers finished working on TAMIGO(v1) and made it ready for deployment, assignment 1 evaluation had already finished. Hence, the TAs started using TAMIGO(v1) for the evaluation of Assignment 2. They generated viva questions and model answers using the question generation module, then provided each student with two questions via a Google Doc. Students shared their screens and turned on their cameras to ensure academic integrity. After students completed writing their responses on the shared Google Doc within a five-minute window, TAs collected the answers and used TAMIGO's answer evaluation module to generate feedback for grading.

\noindent\textbf{Phase 4 - Feedback discussion}
Once the evaluation of Assignment 2 was complete, the TAs and the instructor held discussions to review TAMIGO’s strengths, weaknesses, and suggested improvements based on their experiences with this LLM-based tool.

\noindent\textbf{Phase 5 - Discussion to revamp TAMIGO}
In this phase, the instructor conveyed the feedback collected from the TAs in Phase 4 to the developers. Key improvements included (1) incorporating detailed and structured evaluation criteria such as correctness, errors, and omissions, and (2) presenting the LLM responses to the TAs in plain text for improved readability. Additionally, a new Code Evaluation module was introduced to aid the TAs in assessing student code. This module was designed to generate both a summary and detailed feedback for student code snippets, each corresponding to an essential and critical system functionality, as defined by the instructor.

\noindent\textbf{Phase 6 - Development of TAMIGO(v2)}
Following discussions between the instructor and developers, TAMGIO was revamped by enhancing the Question Generation and Answer Evaluation modules and introducing a new Code Evaluation module.

\noindent \textbf{Phase 7 - Deployment of TAMIGO(v2)}
For Assignment 3, the deployment followed the same evaluation procedure and proctoring as Phase 3. In addition to using the Question Generation and Answer Evaluation modules, TAs utilized the new Code Evaluation module. Students directed TAs to relevant code snippets for each system functionality defined in the rubric. These code snippets were fed into the code evaluation module. The LLM generated detailed evaluations and summaries for each code snippet.

\noindent\textbf{Phase 8 - Final Discussion with TAs}
The instructor held a final discussion with TAs to review their overall experience, discuss the improvements made, and identify any remaining positives, negatives, and further enhancements needed.

\begin{figure*}[htb]
  \includegraphics[width=0.7\textwidth]{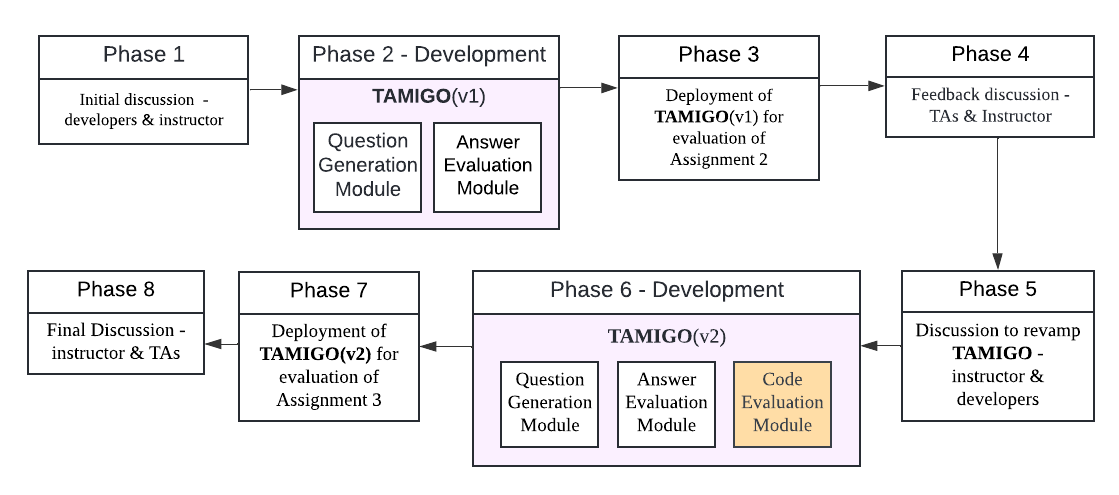}
  \vspace{-2em}
  \caption{Overview of all the phases of development and deployment of TAMIGO: An LLM-based Tool for assisting TAs}
  \label{fig:Overview of Phases}
  \Description[<Phases over time>]{<>}
\end{figure*}

\subsection{Data Collection}
We collected data throughout the various phases of the development and deployment of TAMIGO to comprehensively evaluate its effectiveness and impact.

\noindent\textbf{LLM-Generated Data.} At the end of the evaluation for Assignment 2, TAs were asked to share all the LLM-generated viva questions they used for assessing students, along with the student answers, LLM-generated model answers, and LLM-generated feedback on the student answers. This process was repeated at the end of the evaluation for Assignment 3. Additionally, for Assignment 3, the TAs were also requested to share the LLM-generated summaries and feedback on student code blocks, along with the corresponding student code blocks.

\noindent\textbf{Transcripts of the discussions between the TAs and the instructor.} To gain deeper insights into the evaluation process and the effectiveness of TAMIGO, the instructor conducted two discussion sessions with the TAs following the evaluations of both Assignment 2 and Assignment 3, resulting in a total of four discussion sessions. Each session lasted between 60 and 100 minutes. These sessions were recorded and subsequently transcribed for detailed thematic analysis.
\vspace{-1em}

\subsection{Data Analysis}

\textbf{Analysis of LLM-Generated Data.} This analysis was conducted by two researchers who were domain experts in Distributed Systems and were also conversant with TAMIGO. The analysis for viva assessment and code assessment was performed independently.

\noindent\textbf{(1) Analysis of LLM-generated viva questions, model answers and feedback on the student answers.} An iterative approach was employed by the researchers to refine evaluation metrics. Initially, 20 randomly selected LLM-generated questions were rated by both the researchers using a preliminary set of metrics (suggested by the instructor). Inter-rater reliability (IRR) was calculated using ICC3K \cite{nihGuidelineSelecting} to assess agreement. The researchers discussed their analysis, resolved conflicts and further refined the evaluation metrics. The researchers also discussed their initial analysis with the course instructor and incorporated their feedback. Following this, another set of 20 questions was evaluated. This led to a satisfactory IRR (>0.8). Ultimately, 120 questions (60 per researcher) were evaluated using the finalized metrics. The same methodology was applied to the analysis of model answers and feedback on student answers.


\noindent\textbf{(2) Analysis of LLM-generated summary and feedback on the student code snippets.} 
A similar process was followed for the analysis of LLM-generated summaries and feedback on code snippets. After familiarization with the data, 11 randomly selected code snippets were evaluated by both researchers using initial metrics. Given the satisfactory IRR (ICC3K > 0.75) obtained in the first iteration, a total of 66 code snippets (33 per researcher) were finally assessed using the refined metrics.

The final list of refined metrics used in analysis is defined and described in Section \ref{sec:metrics}.

\noindent\textbf{Analysis of discussion transcripts.} 
Thematic analysis \cite{braun_thematic_analysis} was employed to gain deeper insights into TAs' experiences with TAMIGO. Group discussions between TAs and the instructor were transcribed and analyzed by a single domain expert in distributed systems with in-depth knowledge of TAMIGO. The researcher went through the transcripts multiple times, allowing codes to emerge inductively. Regular discussions with the instructor provided additional context. Ultimately, the identified codes were organized into three overarching themes for each of the three modules of TAMIGO: (1) positive TA experiences with TAMIGO, (2) challenges encountered during TAMIGO use, and (3) suggestions for system enhancement.

\vspace{-1em}

\subsection{Metrics}\label{sec:metrics}
Each LLM-generated response—whether it is a question, model answer, feedback on a student’s answer, code summary, or feedback for a student code snippet—is evaluated using a set of predefined metrics. The ratings for each metric are categorized as follows: (1) \textbf{Binary:} The metric can take only two values, 0 or 1; (2) \textbf{Likert Scale:} The metric can take any integer value between 1 (very poor) and 5 (excellent). The choice between using a binary scale or a Likert scale for any given metric was determined by the researchers at the outset of the analysis, based on their assessment of the potential range of ratings that each metric could appropriately have.

\noindent \textbf{LLM-Generated Questions.}
To evaluate the quality of questions generated by the LLM, we used five metrics: (1) \textbf{Relevance} (Likert Scale) to measure how closely the question relates to assignment topics; (2) \textbf{Clarity} (Likert Scale) to assess if the question is easy to understand; (3) \textbf{Correctness} (Binary) to check if the question is accurate; (4) \textbf{Difficulty} (Likert Scale) to determine if the question is appropriately challenging; and (5) \textbf{Depth} (Likert Scale) to gauge how specific and detailed the question is. 

\noindent \textbf{LLM-Generated Model Answers.}
To evaluate the quality of model answers generated by the LLM, we used four metrics: (1) \textbf{Correctness} (Binary) to ensure accuracy, with any incorrect information leading to a rating of 0; (2) \textbf{Completeness} (Likert Scale) to assess coverage of the question's various aspects; (3) \textbf{Conciseness} (Binary) to check for unnecessary information; and (4) \textbf{Structure} (Binary) to verify proper grammar, spelling, and clarity. 

\noindent \textbf{LLM-Generated Feedback for Student Viva Answers.}
To assess the LLM's feedback on student viva answers, we employed several metrics: (1) \textbf{Correctness} (Binary) to ensure factual accuracy; (2) \textbf{Comprehensiveness}  (Likert Scale), which measures coverage of all relevant points including the detail and clarity in improvement suggestions; 
 (3) \textbf{Constructiveness} (Likert Scale) to determine how helpful the feedback is for student improvement; (4) \textbf{Balanced Nature} (Binary) to check for a mix of criticism and positive reinforcement; (5) \textbf{Consistency} (Binary) to ensure the feedback does not contradict itself and aligns with the model answer; and (6) \textbf{Overwhelming} (Binary) to avoid excessive and unnecessary information. 

\noindent \textbf{LLM-generated feedback for Student Code Snippet.}
To evaluate the quality of the LLM-generated feedback for student code snippets, we used five metrics: (1) \textbf{Rubric Alignment} (Binary) to check if the evaluation follows predefined rubric criteria; (2) \textbf{Constructiveness} (Likert Scale) to measure how helpful the feedback is in providing actionable suggestions; (3) \textbf{Comprehensiveness} (Likert Scale) to assess whether the evaluation covers all key aspects of the code without unnecessary details; (4) \textbf{Balanced Nature} (Binary) to evaluate if the feedback fairly addresses both strengths and weaknesses; and (5) \textbf{Overwhelming} (Binary) to ensure the evaluation is clear and not overly detailed or complex. 

\noindent \textbf{LLM-generated summary for Student Code Snippet.}
To evaluate the quality of code summaries generated by the LLM, we used four metrics: (1) \textbf{Correctness} (Binary) to evaluate if the summary accurately represents the code's functionality and purpose; (2) \textbf{Completeness} (Likert Scale) to measure how well the summary covers all important aspects of the code; (3) \textbf{Conciseness} (Likert Scale) to assess whether the summary is brief and to the point without unnecessary details; and (4) Usefulness (Likert Scale) to determine how helpful the summary is in understanding the code. 

\subsection{Ethical Considerations}
Ethical guidelines were strictly followed to ensure participant privacy and data confidentiality. Data was anonymized, and informed consent was obtained from all participants. They were clearly informed about the research objectives and data usage. Adherence to data protection regulations and institutional policies was maintained throughout the study.

\section{Evaluation \& Results}

\begin{table}
    \centering
    \small
    \begin{tabular}{|>{\centering\arraybackslash}p{0.13\linewidth}|>{\centering\arraybackslash}p{0.13\linewidth}|>{\centering\arraybackslash}p{0.10\linewidth}|>{\centering\arraybackslash}p{0.16\linewidth}|>{\centering\arraybackslash}p{0.12\linewidth}|>{\centering\arraybackslash}p{0.10\linewidth}|} \hline 
         \textbf{Assign - ment}&  \textbf{Relevance} ( 1 to 5)&  \textbf{Clarity} (1 to 5)&  \textbf{Correctness} (0,1)&  \textbf{Difficulty} ( 1 to 5)& \textbf{Depth} (1 to 5)\\ \hline 
         \textbf{\#2}&  4.65 $\pm$ 0.97&  4.37 $\pm$ 0.95&  0.90 $\pm$ 0.30& 3.17 $\pm$ 0.92& 3.22 $\pm$ 0.92\\ \hline 
         \textbf{\#3}& 4.70 $\pm$ 0.81& 4.87 $\pm$ 0.43& 1.00 $\pm$ 0.00& 3.48 $\pm$ 0.79& 3.47 $\pm$ 0.81\\ \hline
    \end{tabular}
    \caption{Avg. Ratings for Questions Generated by TAMGIO.}
    \label{table:question_ratings}
\end{table}

\noindent \textbf{Analysis of Questions Generated by LLMs.} Table \ref{table:question_ratings} demonstrates that the questions generated by the LLM for both assignments exhibited high relevance and clarity, with mean scores exceeding 4.3 and low standard deviations, indicating consistency. The correctness score improved from 0.9 in Assignment 2 to a perfect 1 in Assignment 3, reflecting enhanced accuracy. The difficulty and depth scores were moderate for both assignments, suggesting that the questions were appropriately challenging and specific, without being overly simplistic or superficial. Overall, the metrics indicate strong performance across all the metrics for both assignments. 


\begin{table}[h]
    \centering
    \small
    \begin{tabular}{|>{\centering\arraybackslash}p{0.13\linewidth}|>{\centering\arraybackslash}p{0.165\linewidth}|>{\centering\arraybackslash}p{0.195\linewidth}|>{\centering\arraybackslash}p{0.17\linewidth}|>{\centering\arraybackslash}p{0.125\linewidth}|} \hline 
         \textbf{Assign - ment}&  \textbf{Correctness} (0, 1)&  \textbf{Completeness} (1 to 5)&  \textbf{Conciseness} (0,1)&  \textbf{Structure} (0, 1)\\ \hline 
         \textbf{\#2}&  0.7 $\pm$ 0.46&  3.58 $\pm$ 1.41&  0.63 $\pm$ 0.48& 0.92 $\pm$ 0.28\\ \hline 
         \textbf{\#3}& 0.73 $\pm$ 0.43& 4.37 $\pm$ 0.86& 0.8 $\pm$ 0.40& 1 $\pm$ 0\\ \hline
    \end{tabular}
    \caption{Ratings for Model Answers Generated by TAMIGO.}
    \label{table:model_answer_ratings}
    \vspace{-2em}
\end{table}

\noindent \textbf{Analysis of Model Answers Generated by LLMs.} Table \ref{table:model_answer_ratings} shows that the model answers showed improvement from Assignment 2 to Assignment 3, with correctness increasing from 0.7 to 0.73 and completeness significantly rising from 3.58 to 4.37, indicating better coverage of question aspects. Conciseness also improved, from 0.63 to 0.8, showing reduced unnecessary information. Structure was consistently high, reaching a perfect 1 in Assignment 3. These metrics indicate good performance, with significant enhancements in completeness and conciseness for Assignment 3. The improvements in metrics for Assignment 3 may be attributed to the fact that Assignment 3 was on a relatively standard and simpler topic (MapReduce and K-means) as compared to Assignment 2 which was on a tweaked version of a standard consensus algorithm (Raft). 

\begin{table}[h]
    \centering
    \footnotesize
    \begin{tabular}{|>{\centering\arraybackslash}p{0.08\linewidth}|>{\centering\arraybackslash}p{0.10\linewidth}|>{\centering\arraybackslash}p{0.12\linewidth}|>{\centering\arraybackslash}p{0.12\linewidth}|>{\centering\arraybackslash}p{0.12\linewidth}|>{\centering\arraybackslash}p{0.09\linewidth}|>
    {\centering\arraybackslash}p{0.10\linewidth}|} \hline 
         \textbf{Assign - ment}&  \textbf{Correct - ness} (0, 1)&  \textbf{Construct - iveness} (1 to 5)&  \textbf{Compre - hensive - ness} (1 to 5)&  \textbf{Balanced Nature} (0, 1)& \textbf{Consis - tency} (0, 1)& \textbf{Over - whel - ming} (0, 1)\\ \hline 
         \textbf{\#2}&  0.73 $\pm$ 0.44&  3.67 $\pm$ 1.16&  3.47 $\pm$ 0.98& 0.83 $\pm$ 0.37& 0.80 $\pm$ 0.40& 0.45 $\pm$ 0.50\\ \hline 
         \textbf{\#3}& 0.55 $\pm$ 0.50& 3.43 $\pm$ 1.18& 3.40 $\pm$ 1.16& 0.98 $\pm$ 0.12& 0.75 $\pm$ 0.43& 0.20 $\pm$ 0.40\\ \hline
    \end{tabular}
    \centering\caption{Ratings for TAMIGO Feedback for Student Answers.}
    \label{table:viva_evaluation_ratings}
    \vspace{-1em}
\end{table}
\vspace{-1em}
\noindent \textbf{Analysis of LLM-Generated Feedback for Student Answers} Table \ref{table:viva_evaluation_ratings} shows mixed results for the feedback generated by the LLMs in the context of evaluating student answers. While the correctness of the feedback provided by the LLM was decently good for assignment 2, it decreased to 0.55 for assignment 3. This was due to the hallucinations for accurate student answers when there was actually no error in the student answer. The LLM was identifying incorrect errors as it was specifically asked to identify the omissions and errors in the student answers. Constructiveness, comprehensiveness and consistency were good, suggesting consistent quality in providing helpful and thorough feedback for both the assignments. Metrics for balanced nature as well as overwhelming nature improved significantly from assignment 2 to assignment 3, with balanced nature rising from 0.83 to 0.98 and overwhelming dropping from 0.45 to 0.2, indicating more balanced and concise feedback. This was due to the fact that the LLM was specifically asked to generate structured feedback based on correctness, omissions, errors and overall evaluation.

\begin{table}[h]
    \centering
    \resizebox{\columnwidth}{!}{%
    \renewcommand{\arraystretch}{1.6} 
    \newcolumntype{Y}{>{\centering\arraybackslash}p{2.76cm}} 
    \fontsize{15}{14}\selectfont
        \begin{tabular}{|Y|Y|Y|Y|}
            \hline
            \textbf{Correctness (0, 1)} & \textbf{Complete-ness (1 to 5)} & \textbf{Conciseness (1, 5)} & \textbf{Usefulness (1, 5)} \\
            \hline
            0.98 $\pm$ 0.13 & 4.61 $\pm$ 0.76 & 3.86 $\pm$ 0.99 & 3.9 $\pm$ 0.95 \\
            \hline
        \end{tabular}
    }
    \caption{Ratings for the LLM-Generated Code Summaries for Student Code Snippents in Assignment 3.}
    \label{table:code_summary_ratings}
    \vspace{-2em}
\end{table}

\noindent \textbf{Analysis of LLM-Generated Code Summaries.} Table \ref{table:code_summary_ratings} shows that the code summaries generated by the LLM in Assignment 3 demonstrated high quality. Correctness was nearly perfect at 0.98, indicating accurate representations of code functionality. Completeness had a strong mean of 4.61, showing thorough coverage of important aspects. Conciseness was good at 3.86, reflecting brief and focused summaries. Usefulness scored 3.9, suggesting the summaries were helpful for understanding the code. Overall, these metrics indicate that the LLM produced accurate, comprehensive, and useful code summaries.

\begin{table}[h]
    \centering
    \resizebox{\columnwidth}{!}{%
    \renewcommand{\arraystretch}{1.6} 
    \newcolumntype{Y}{>{\centering\arraybackslash}p{2.76cm}} 
    \fontsize{15}{14}\selectfont
        \begin{tabular}{|Y|Y|Y|Y|Y|}
            \hline
            \textbf{Rubric - Alignment (0, 1)} & \textbf{Construct- iveness (1 to 5)} & \textbf{Comprehen- siveness (1 to 5)} & \textbf{Balanced Nature (0, 1)} & \textbf{Over- whelming (0, 1)} \\
            \hline
            0.68 $\pm$ 0.47 & 3.78 $\pm$ 0.94 & 3.78 $\pm$ 0.94 & 0.95 $\pm$ 0.22 & 0.20 $\pm$ 0.40 \\
            \hline
        \end{tabular}
    }
    \caption{Ratings for the LLM-Generated Feedback for Student Code Snippets in Assignment 3.}
    \label{table:code_evaluation_ratings}
    \vspace{-2em}
\end{table}

\noindent \textbf{Analysis of LLM-Generated Feedback for Student Code.} Table \ref{table:code_evaluation_ratings} shows encouraging results for the feedback generated by the LLM for student code snippets in Assignment 3. Rubric Alignment was moderate at 0.68, suggesting some adherence to predefined criteria. Both Constructiveness and Comprehensiveness had good scores of 3.78, indicating helpful and thorough feedback. Balanced Nature was high at 0.95, reflecting fair feedback on strengths and weaknesses. Overwhelming was 0.2, showing the evaluations were easy to understand and not too much to go through. Overall, the evaluations were constructive, comprehensive, balanced and not overwhelming, though there is room for improvement in following rubric criteria.
\begin{table}[]
\centering
\begin{tabular}{l|l|l|l|}
    \cline{2-4}
                                       & QG     & AE     & CE \\ \hline
    \multicolumn{1}{|l|}{Assignment 2} & 3.785 $\pm$ 0.42 & 3.57 $\pm$ 0.64 &  NA \\ \hline
    \multicolumn{1}{|l|}{Assignment 3} & 3.78 $\pm$ 0.89  & 3.78 $\pm$ 0.69 & 3.64 $\pm$ 0.74\\ \hline
    \end{tabular}
    \caption{Average Ratings for Overall Experience of TAs for different modules of TAMIGO: Question Generation(QG), Answer Evaluation(AE) and Code Evaluation (CE)}
    \label{table:overall_ratings}
    \vspace{-1em}
\end{table}
\noindent \textbf{Overall Experience of TAs with TAMIGO:}
Table \ref{table:overall_ratings} presents the average ratings for TAs' overall experience with different TAMIGO modules: Question Generation (QG), Answer Evaluation (AE), and Code Evaluation (CE). Similar ratings were received for Assignment 2 and Assignment 3.
Overall, the consistency in ratings across assignments indicates stable performance of TAMIGO and room for further improvement and enhancement.


\noindent \textbf{Analysis of Group Discussions:} 
The Overall experience for each module was evaluated using Likert scale (1 (very poor) to 5(excellent) )

\noindent \textbf{(1) Question Generation Module:} Participants highlighted both positive aspects, such as question relevance and clarity, and negatives, including repetitive or overly generic questions. Suggestions for improvement centered on diversifying question types and enhancing specificity.

\noindent\textbf{(2) Answer Evaluation Module:} The evaluation of student answers revealed mixed sentiments. Positive feedback noted clearer delineation between correct and incorrect responses, yet criticisms centered on occasional inaccuracies and the need for more comprehensive model answers. Suggestions at the end of assignment 2 discussion included refining the model answer criteria and providing more detailed feedback to aid student comprehension. This was incorporated before the evaluation of assignment 3 and was appreciated by the TAs during the feedback session at the end of assignment 3 evaluation.

\noindent\textbf{(2) Code Evaluation Module:} 
Positive feedback acknowledged coverage of code functionalities, but concerns were raised about incomplete or unreliable outputs. Recommendations focused on enhancing the accuracy and reliability of generated code summaries and evaluations by providing more context about student code to the LLM.

\noindent\textbf{(4) Impact on TA Workload:} While some TAs noted a decrease in workload due to streamlined LLM-based evaluation, others highlighted increased complexity, particularly with the addition of code evaluation tasks. Suggestions included streamlining UI and automating some of the manual tasks which the TAs were doing while interacting with TAMIGO.
\vspace{-1em}
\section{Limitations}

Despite the promising results observed in our study, there were several limitations of TAMIGO. Firstly, TAMIGO does not provide numerical grades, which may limit its applicability in contexts where precise grading is essential. Secondly, the system generated substantial amounts of feedback data from student interactions, yet due to resource constraints, not all data could be fully analyzed, potentially missing deeper insights. Moreover, while TAMIGO aimed to streamline the workflow for TAs, there were occasional challenges reported regarding its seamless integration into their existing processes, suggesting room for improvement in user interface and usability. Additionally, the effectiveness of TAMIGO's feedback was influenced by the expertise available during its development phase, highlighting the need for more advanced strategies and ongoing refinement to enhance its accuracy and utility. Furthermore, TAMIGO did not evaluate student perception using AI-based evaluators, which could have provided insights into how students perceive and respond to AI-generated feedback.

\vspace{-1em}
\section{Conclusion and future work}
This paper demonstrates the potential of Large Language Models (LLMs) in enhancing the efficiency and effectiveness of teaching assistants (TAs) in advanced computing courses. Through the deployment of TAMIGO, an LLM-based system, we facilitated viva and code assessments in a distributed systems class, leveraging GPT-3.5-Turbo to generate questions and provide feedback. Our analysis shows that while LLMs excel in creating contextually relevant and comprehensive questions and feedback, there are challenges to address, particularly concerning the accuracy and alignment of feedback with evaluation rubrics. Despite occasional hallucinations, the consistent and constructive nature of LLM-generated feedback indicates a promising role for LLMs in educational settings. Future research should focus on refining these systems to mitigate inaccuracies and further align their output with instructor-provided rubrics. Our findings contribute valuable insights into the integration of AI tools in academia, paving the way for more effective and balanced educational support mechanisms.

\bibliographystyle{ACM-Reference-Format}
\bibliography{chatgpt-ref}

\appendix

\end{document}